\documentclass[showpacs,twocolumn,superscriptaddress]{revtex4-2}

\usepackage{hyperref}
\hypersetup{
  colorlinks=true,        
  linkcolor=blue,         
  citecolor=cyan,         
}

\def\be{\begin{equation}}
\def\ee{\end{equation}}
\def\bestar{\begin{equation*}}
\def\eestar{\end{equation*}}

 \usepackage{graphicx}
\newcommand{\lb}{\label}

\usepackage{amsmath}

\newcommand{\bw}{\begin{widetext}}
\newcommand{\ew}{\end{widetext}}

\newcommand{\ba}{\begin{eqnarray}}
	\newcommand{\ea}{\end{eqnarray}}
\newcommand{\bal}{\begin{align}}
	\newcommand{\eal}{\end{align}}

\usepackage[utf8]{inputenc}
\usepackage{amssymb}
\usepackage{amsmath}
\usepackage{amsfonts}
\usepackage{graphicx, float}
\usepackage{graphicx, epsfig}
\usepackage{color}
\usepackage{enumerate}
\newcommand{\beq}{\begin{equation}}
	\newcommand{\eeq}{\end{equation}}
\newcommand{\bqn}{\begin{eqnarray}}
	\newcommand{\eqn}{\end{eqnarray}}

\newcommand{\EpMod}{\ensuremath{ \mathfrak{T} }}

\newcommand{\der}{\ensuremath{ d }}

\def\ber{\begin{eqnarray}}
\def\eer{\end{eqnarray}}

\begin{document}

\title{Superradiance and stability of rotating charged black holes in T-duality}
\author{Sohan Kumar Jha}
\email{sohan00slg@gmail.com}
\affiliation{Department of Physics, Chandernagore College, Chandernagore, Hooghly, West
Bengal, India}

\author{Kimet Jusufi}
\email{kimet.jusufi@unite.edu.mk}
\affiliation{Physics Department, State University of Tetovo, Ilinden Street nn, 1200, Tetovo, North Macedonia}

\begin{abstract}
We investigate the shadow images, the relation between Quasinormal Modes (QNMs) and the shadow radius, and the superradiance effect observed in the context of a rotating charged black hole under T-duality. Our investigation places particular emphasis on two key parameters: the electric charge denoted as $Q$ and the quantum deformed parameter represented by the zero-point length, $l_0$. Our findings reveal a distinct pattern: as the electric charge increases, the shadow radius experiences a consistent decrease. Intriguingly, when considering the quantum deformed parameter, we find a noteworthy phenomenon—a reflecting point. Specifically, we illustrate that the shadow radius initially increases with an increase in $l_0$ and subsequently decreases. Further analysis involves the computation of eikonal equatorial and polar QNMs, where a similar reflecting point emerges upon varying $l_0$. This establishes the inverse correlation between QNMs and shadow radius within our research framework. 
Our investigation into the effects of Q and $l_0$ on superradiance reveals that the amplification factor initially increases with $Q$ and $l_0$ and then starts decreasing. Moreover, the rotating black holes in T-duality allows superradiance scattering for a wider range of frequency than Kerr black holes, making black holes in T-duality brighter than the Kerr black holes. We also delve into the stability of the combined system of the rotating black hole and scalar field with the help of black hole bomb mechanism. It provides a window to observe the impact of parameters Q and $l_0$ on the stability. It shows that the combined system is stable for a wider regime for the Kerr black hole. 

\end{abstract}

\maketitle

\renewcommand{\thefootnote}{\arabic{footnote}} \setcounter{footnote}{0}

\section{Introduction}
Numerous contemporary astrophysical observatories provide strong evidence supporting the presence of supermassive black holes (SMBHs) at the cores of giant elliptical and spiral galaxies. One illustrative instance is the observation indicating the existence of a supermassive black hole with a mass equivalent to four million solar masses at the heart of our Milky Way galaxy \cite{m87,m871,EHT2022-1,EHT2022-2,EHT2022-3}. These supermassive  black holes are characterized by their masses and spin parameters.

According to Einstein's theory of relativity, black holes typically contain a spacetime singularity at their center and an event horizon where gravity is so intense that nothing, not even electromagnetic radiation or light, can escape. Due to the external spacetime geometry, a black hole (BH) can capture light from nearby stars or accretion disks into bound orbits. In simpler terms, each black hole is characterized by a photon sphere—a collection of light rays orbiting the BH. It's noteworthy that the orbit of light is considered unstable if the photon can either fall into the black hole or escape to infinity.

Another compelling evidence for the existence of supermassive black hole comes from the first shadow images at the center of the M87 galaxy that was captured by the Event Horizon Telescope (EHT) collaboration \cite{2,3}. Additionally, the detection of gravitational waves by LIGO contributes to our understanding \cite{4}. Intriguingly, by monitoring the motion of stars in the Galactic Center and collecting precise measurements of their orbital motion, we can delve into the nature of black holes and the surrounding spacetime.

Superradiance is a radiation enhancement process where it is possible to extract energy from the black hole under certain conditions. Penrose first conceptualized the exciting idea of energy extraction from black holes in seminal works \cite{penrose1, penrose2}. Later, Misner, Zel'dovich, and Bekenstein independently derived the essential condition, $\omega<m\Omega$, for superradiance where $\omega$, m, and $\Omega$ are the frequency of the incident wave, the azimuthal quantum number with respect to the rotational axis, and the angular velocity of the rotating body, respectively [\citenum{misner}-\citenum{bekenstein2}]. The same inequality holds if we consider the superradiance scattering of electromagnetic and gravitational waves \cite{teukolsky}. It was various investigations in the field of superradiance that played an instrumental role in the discovery of black hole evaporation \cite{hawking2}. Studies such as \cite{richartz, cardoso} may give an impression of superradiance's sole dependence on the horizon boundary condition. However, numerous studies reveal that the requisite dissipation essential for the superradiance scattering is provided by the ergoregion [\citenum{ge}-\citenum{page}]. It is thus possible to observe superradiance in the case of horizonless stars [\citenum{ge}-\citenum{kg}]. Instabilities induced by superradiance in the background spacetime may lead to a hairy black hole solution, thus providing an excellent window to investigate the no-hair theorem \cite{khodadi}. Superradiance provides an avenue to probe the stability of black holes and explore the gravitational collapse in confining geometries \cite{ads0}.

In this manuscript, our focus is on investigating both the superradiance effect and the Quasinormal Modes (QNMs) of a charged black hole identified in T-duality \cite{Gaete:2022ukm}. This particular black hole spacetime is characterized by parameters such as mass, spin, electric charge, and the quantum-deformed parameter, also known as the zero-point length parameter. The geometry of this spacetime is asymptotically flat, the solution is regular at the origin, and serves as a generalization of the Kerr–Newman solution. In the realm of astrophysical black holes, the primary influencers are anticipated to be the mass, spin, and electric charge. However, for smaller black holes and when exploring the interior structure, quantum effects may play a significant role. Our exploration delves into understanding these phenomena in the context of the T-duality framework. 

The paper is organized as follows. In Section II, we review the charged black hole solution in T-duality. In Section III we investigate the shadow images and the relation between QNMs and shadow radius.  In Section IV, we discuss the superradiance effect and the stability of the black hole. Finally, in Section V, we comment on our findings. 

\section{The BH solution}

In this section, we shall review the exact solution for a charged black hole in T-duality recently found in \cite{Gaete:2022ukm}. In this setup, one starts by assuming the general solution to be static and spherically symmetric in the form
\begin{equation}
\label{eq:lineElem}
\der s^2
=- f(r)\, dt^2 +\frac{dr^2}{g(r)} +r^2(d\theta^2+\sin^2\theta d\phi^2)
\end{equation}
with
\begin{equation}
\begin{split}
g(r)
&=1 - \frac{2 m(r)}{r}
\end{split}
\end{equation}
where the mass profile is given by
\begin{equation}
m(r)=-4\pi\int_0^r dr^\prime (r^\prime)^2\ \EpMod_0^0(r^\prime).
\end{equation}

In addition, it is assumed that the effective Einstein field equation holds  \cite{Gaete:2022ukm}
\begin{eqnarray}
&& R_{\mu\nu}-\frac{1}{2}g_{\mu\nu} R= \kappa \left(\mathfrak{T}_{\mu\nu}^\mathrm{bare}+\mathfrak{T}_{\mu\nu}^\mathrm{em}\right)
\end{eqnarray}
where $\kappa=8 \pi$, along the modified energy-momentum tensor.  Namely we have a quantum-modified source as well as the modified electromagnetic field
\begin{equation}
 \EpMod_{\mu \nu}^\mathrm{em}=\frac{1}{4 \pi}\left(F_{\mu \sigma}{F_{\nu}}^{\sigma}    -\frac{1}{4}g_{\mu \nu}F_{\rho \sigma}F^{\rho \sigma}\right).
 \label{eq:emtem}
\end{equation}

In T-duality, it has been shown that the gravitational potential is modified and given by the non-singular expression \cite{Gaete:2022ukm}
\begin{equation}
V(r)=-\frac{m_0}{\sqrt{r^2 +l_0^2}}.
\end{equation}

From the Newton-Poisson equation, one finds
\begin{equation}
\rho_\mathrm{bare}(r)=\frac{1}{4\pi}\Delta V(r)= \frac{3 l_0^2 m_0}{4 \pi {\left(r^2 +l_0^2\right)}^{5/2}}.
\end{equation}
This expression can be viewed as a generalization of the conventional Newton-Poisson equation, where the source is usually given in terms of the Dirac delta function. In order to obtain the full solution, one has to consider the effect of the electromagnetic field using the modified potential \cite{Gaete:2022une}
\begin{equation}
V_{em}(r)=-\frac{Q}{\sqrt{r^2 + l_0^2}},
\end{equation}
From where one has
\begin{equation}
 \EpMod_{0}^{0\ \mathrm{em}}=\frac{Q^2 r^2}{8 \pi (r^2+l_0^2)^3}.
 \label{eq:estaticen}
\end{equation}
As can be seen this expression vanishes at the origin, and it is decaying as $\sim 1/r^4$ at the asymptotic region at infinity. With these results in hand, one can work out the following exact solution \cite{Gaete:2022ukm}
\begin{eqnarray}\notag
  f(r)&=&g(r)=1-\frac{2m_0 r^2}{\left(r^2+l_0^2\right)^{3/2}}+\frac{5 Q^2 r^2}{8 (r^2+l_0^2)^2}\\
&+&\frac{3 Q^2 l_0^2}{8 (r^2+l_0^2)^2}-\frac{3 Q^2}{8 l_0\,r }\arctan\left(\frac{r}{l_0}\right).
\end{eqnarray}
This is a regular spacetime with the ADM mass $M$ given by (see for details \cite{Gaete:2022ukm})
\begin{equation}
M=m_0+\frac{3\pi Q^2}{32\ l_0}.
\label{eq:admmass}
\end{equation} 
Surprisingly, one finds an intriguing revelation: the mass encompasses an additional term directly proportional to the regularized self-energy of the electrostatic field. In the realm of general relativity, this particular term is typically disregarded, as it is considered integral to the curvature singularity.

It follows that one can employ the expression \eqref{eq:admmass} to formulate the metric coefficient in its ultimate representation:
\begin{equation}
f(r)=1-\frac{2Mr^2}{\left(r^2+l_0^2\right)^{3/2}}+ \frac{Q^2 \ r^2}{\left(r^2+l_0^2\right)^{2}}F(r)
\label{eq:lineelfinal}
\end{equation}
where
\begin{eqnarray}
F(r)&=&\frac{5}{8}+\frac{3l_0^2}{8r^2}+\frac{3\pi}{16l_0}\left(r^2+l_0^2\right)^{\frac{1}{2}}\\\notag
&\times& \left\{1-\frac{\left(r^2+l_0^2\right)^{\frac{3}{2}}}{r^3}\left[\frac{2}{\pi}\arctan\left(\frac{r}{l_0}\right)\right]\right\}.
\end{eqnarray}

 \begin{figure*}[ht!]
		\centering
	\includegraphics[scale=0.6]{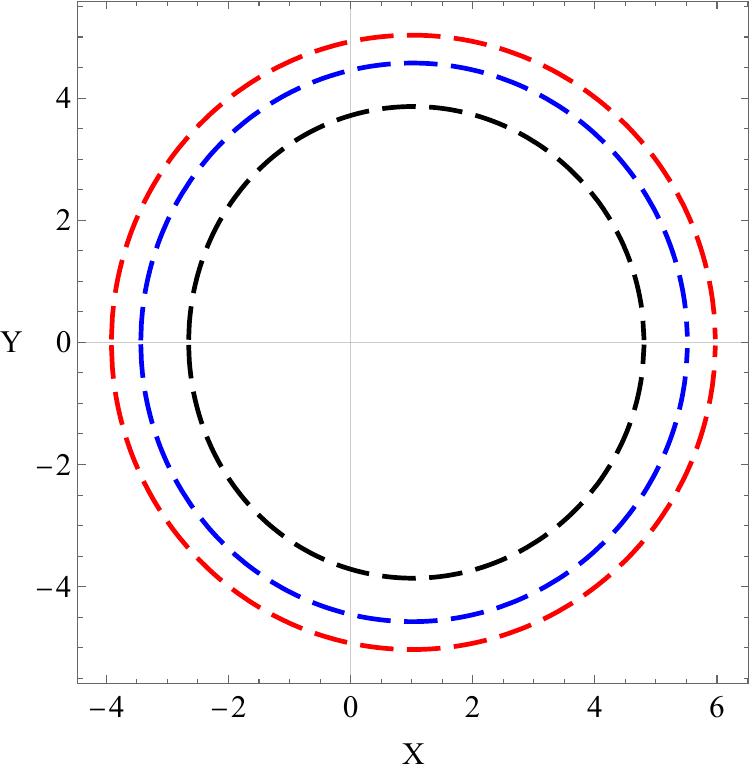}
	\includegraphics[scale=0.6]{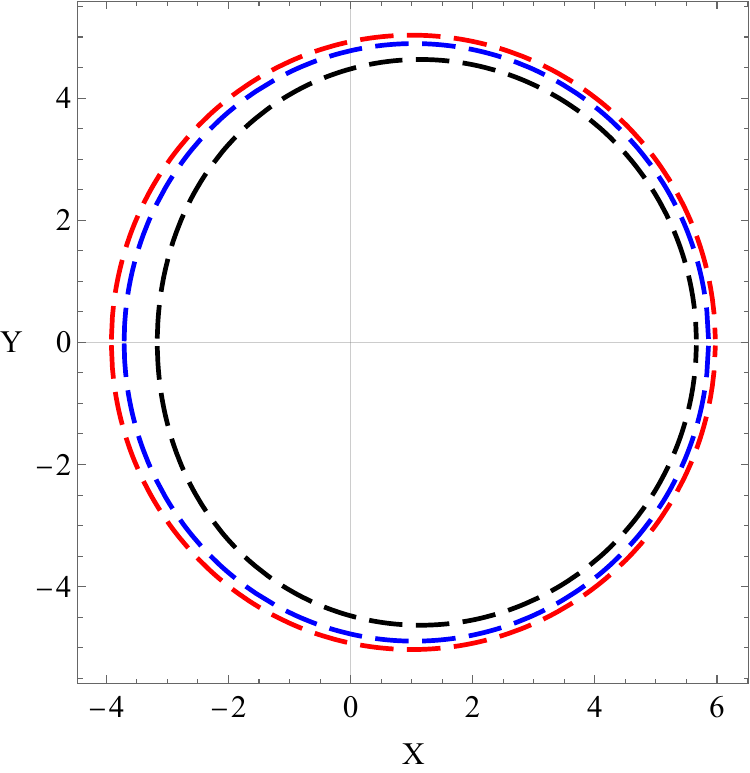}
	\caption{(Left panel) Shadow of the rotating charged black hole for a fixed $l_0=0.1$ (red curve $Q=0.1$, blue curve $Q=0.2$ and black curve $Q=0.3$) and $M=1$, $a=0.5$ and viewing angle $\theta_0=\pi/2$. (Right panel) Shadow of the rotating charged black hole for a fixed $Q=0.1$ (red curve $l_0=0.1$, blue curve $l_0=0.3$ and black curve $Q=0.5$) and $M=1$, $a=0.5$ and viewing angle $\theta_0=\pi/2$.   }\label{shadowq}
	\end{figure*}
 
To get the rotating solution,  we can utilize the non-complexification procedure and we can drop the complexification of the radial coordinate \cite{Azreg-Ainou:2014pra}. Finally, in terms of the Kerr-like coordinates, one can show that \cite{Gaete:2022ukm}
	\begin{eqnarray}\notag
		ds^2 &=& -\frac{\Delta}{\Sigma}(dt-a\sin^2\theta d\phi)^2+\frac{\Sigma}{\Delta}dr^2+\Sigma\,d\theta^2 \\
		\label{metric}	&+&\frac{\sin^2\theta}{\Sigma}[a dt-(r^2+a^2)d\phi]^2, 
		\label{rotating}
	\end{eqnarray}
	with
	\begin{eqnarray}\notag
    \Delta(r)&=&r^2-\frac{2M r^4}{\left(r^2+l_0^2\right)^{3/2}}+\frac{Q^2 r^4}{(r^2+l_0^2)^2} F(r)+a^2,
\end{eqnarray}
along with
	\begin{eqnarray}
		\Sigma(r,\theta)&=&r^2+a^2\cos^2\theta,
	\end{eqnarray}
	where we note that $a=L/M$ is the specific angular momentum, and $M$ is the ADM mass of the black hole.

	\section{Black hole shadow and the correspondence with QNMs \label{secqnm}}
	In this section, our focus shifts to the study of the evolution of photons in the vicinity of the spinning black hole in T-duality. We can start the analysis by using the Hamilton-Jacobi equation
	\bqn
	\frac{\partial S}{\partial \lambda}=-\frac{1}{2}g^{\mu\nu}\frac{\partial S}{\partial x^\mu}\frac{\partial S}{\partial x^\nu},
	\eqn
	where $\lambda$ is an affine parameter. Next, we can write the Jacobi action $S$ in the standard form
	\begin{equation}
		S=\frac{1}{2}m_0^2\lambda-Et+L_z\phi+S_r(r)+S_\theta(\theta),
	\end{equation}
	where we shall use the fact that for photons, the rest mass is zero ($m_0=0$). Furthermore $E$ and $L_z$ are the energy and angular momentum of the photon, respectively. The functions $S_r(r)$ and $S_\theta(\theta)$ are introduced, with $S_r$ being dependent solely on $r$, and $S_\theta$ dependent solely on $\theta$.

  \begin{figure*}[ht!]
		\centering
	\includegraphics[scale=0.65]{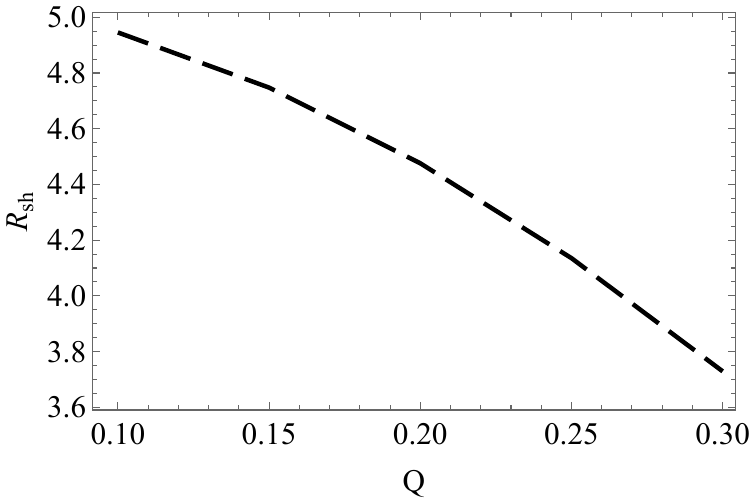}
	\includegraphics[scale=0.65]{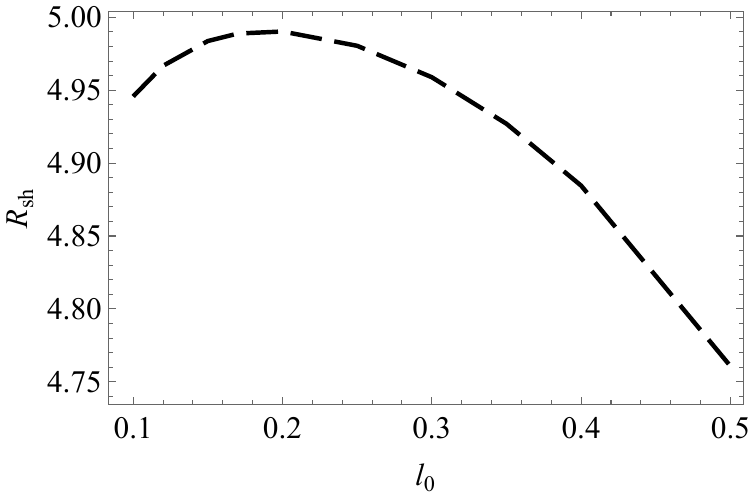}
	\caption{(Left panel) The plot of the typical shadow radius with the viewing angle $\theta_0=\pi/2$ as a function of $Q$. We have fixed $l_0=0.1$ and $M=1$, $a=0.5$ and viewing angle $\theta_0=\pi/2$. (Right panel) The plot of the typical shadow radius with the viewing angle $\theta_0=\pi/2$ as a function of $l_0$. We have fixed $Q=0.1$, $M=1$, $a=0.5$ and viewing angle $\theta_0=\pi/2$, respectively.   }\label{shadowRADIUS}
	\end{figure*}
 To this end, by substituting the Jacobi action into the Hamilton-Jacobi equation, we obtain
	\bqn
	S_r(r)&=&\int^r\frac{\sqrt{R(r)}}{\Delta}dr,\\
	S_\theta(\theta)&=&\int^\theta\sqrt{\Theta(\theta)}d\theta,
	\eqn
	where
	\begin{equation}
		R(r) =[(r^2+a^2)E-aL_z]^2-\Delta[ \mathcal{K}+(L_z-aE)^2],
  \end{equation}
  \begin{equation}
		\Theta(\theta)= \mathcal{K}+(a^2E^2-L^2_z\csc^2\theta)\cos^2\theta,
	\end{equation}
	where $\mathcal{K}$ is the Carter constant. The Hamilton-Jacobi equation leads to the derivation of the following four equations of motion governing the trajectory of photons:
	\bqn \notag
	\rho^2\frac{dt}{d\lambda} &=&a(L_z-aE\sin^2\theta)+\frac{r^2+a^2}{\Delta}[(r^2+a^2)E -aL_z], \lb{YYY} \\\notag
	\rho^2\frac{d\phi}{d\lambda} &=&\frac{L_z}{\sin^2\theta}-aE+\frac{a}{\Delta}[(R^2+a^2)E-aL_z],\\\notag
	\rho^2\frac{dr}{d\lambda} &=&\sqrt{R(r)},\\
	\rho^2\frac{d\theta}{d\lambda}&=&\sqrt{\Theta(\theta)}.\lb{XXX}
	\eqn
	Let us define the two impact parameters
	\bqn
	\xi=\frac{L_z}{E},\qquad \eta=\frac{\mathcal{K}}{E^2},
	\eqn
	which will be used to investigate the geometric configuration of the black hole shadow. To achieve this, we employ the unstable condition associated with circular geodesics given by
	\bqn
	R(r)=0,\qquad \frac{dR(r)}{dr}=0.\lb{ZZZ}
	\eqn

In the general scenario, photons emitted by a light source undergo deflection as they traverse near the black hole due to gravitational lensing effects. Consequently, some photons reach a distant observer after being deflected by the black hole, while others directly plunge into the black hole. The photons unable to escape from the black hole collectively shape the observable shadow of the black hole in the observer's sky. The boundary of this shadow outlines the apparent silhouette of the black hole. To examine the characteristics of the shadow, we adopt celestial coordinates as defined below:
	\begin{eqnarray}
		X&=&-\xi \csc{\theta_0},\\
		Y&=&\pm\sqrt{\eta+a^2 \cos^2\theta_0-\xi^2 \cot^2\theta_0}.
	\end{eqnarray}
	where  \cite{Shaikh:2019fpu}
	\begin{eqnarray}
		\xi(r)&=&\frac{\mathcal{X}_\text{0}\Delta'_\text{0}-2\Delta_\text{0}\mathcal{X}'_\text{0}}{a\Delta'_\text{0}},\\
		\label{eq:xi}
		\eta(r)&=&\frac{4a^2\mathcal{X}'^2_\text{0}\Delta_\text{0}-\left[\left(\mathcal{X}_\text{0}-a^2\right)\Delta'_\text{0}-2\mathcal{X}'_\text{0}\Delta_\text{0} \right]^2}{a^2\Delta'^2_\text{0}}.
		\label{eq:eta}
	\end{eqnarray}
	with $\mathcal{X}=r^2+a^2$. Here, we note that the subscript ``$\text{0}$" indicates the above equations should be evaluated at $r=r_\text{0}$, which is a solution to~\eqref{ZZZ}. In Fig. 1, we give the shadow images of the black holes by varying the electric charge (left panel) and by varying the $l_0$ (right panel). We observe that by keeping a fixed value for $l_0$, the shadow radius decreases with the increase of the electric charge. We can see more clearly this fact from the plot of the typical shadow radius in Fig. 2 (left panel). On the other hand, the situation is more interesting when we fix the electric charge $Q$, and we increase the value of $l_0$. As can be seen from Fig. 2 (right panel), initially, the typical shadow radius increases very slowly with the increase of $l_0$, and then around some critical value around $l_0 \sim 0.2$, the shadow radius reaches the maximal value. With a further increase of $l_0$, the shadow radius decreases faster.

 \subsection{The correspondence between shadow radius and  eikonal QNMs}
 In this section, we elaborate the relation between the QNMs, and the shadow radius (see, \cite{cardoso,stefanov,J1,Jusufi:2019ltj,Konoplya:2017wot,Yang:2012he,Y,c1,mash,Dolan}). It is well-known that QNMs modes  are associated with the ringdown phase of a black hole, can be expressed in terms of their real and imaginary parts as $\omega=\omega_{\Re}- i \omega_{\Im}$. An intriguing observation was shown in Ref. \cite{Yang:2012he}, where it was argued that the QNM frequency of a Kerr black hole in the eikonal limit is given by:
	\begin{eqnarray}
		\omega_{QNM}=(l+\frac{1}{2})\Omega_R-i \gamma_L \left(n+\frac{1}{2}\right)
	\end{eqnarray}
	with 
	\begin{eqnarray}
		\Omega_R=\Omega_{\theta}+\frac{m}{l+\frac{1}{2}}\Omega_{prec},
	\end{eqnarray}
	where $\Omega_{\theta}$ represents the orbital frequency in the polar direction, $\Omega_{\text{prec}}$ corresponds to the Lense-Thirring precession frequency of the orbit plane, $\gamma_L$ stands for the Lyapunov exponent of the orbit, and $n$ denotes the overtone number. In the upcoming discussion, we would like to study two specific modes, equatorial and polar modes, along with the effect of electric charge and zero point length.

         \subsection{Viewing angle: $\theta_0=\pi/2 $}
	
	In the general case of a rotating black hole, there is no closed form expression for the shadow radius. In what follows, we shall consider the case of a viewing angle:  $\theta_0=\pi/2 $. For simplicity, we shall also consider equatorial orbit, which can be used to compute the typical shadow radius. We are also going to use the following fact that the Lense-Thiring precession frequency is related to the orbital frequency and Keplerian frequency, namely, the lense–thirring precession frequency for prograde orbits in the limit of a small perturbation with respect to the equatorial plane is as follows
	\begin{eqnarray}
		\Omega_{prec}=\pm \Omega_{\phi}\mp \Omega_{\theta}\label{lt}
	\end{eqnarray}
	where 
	\begin{eqnarray}\label{Omegaf}
		\Omega_{\phi}=\frac{-\partial _r g_{t \phi }\pm \sqrt{\left(\partial _r g_{t \phi }\right)^2-(\partial _r g_{t t})( \partial _r g_{\phi  \phi })}}{\partial _r g_{\phi
				\phi }}.
	\end{eqnarray}
	
	In the case of spinning metric in Ref. \cite{J1} it was studied the case $m \pm l$, while in Ref. \cite{Y} it was argued that 
	\begin{eqnarray}
		\mathcal{K}+L_z^2 \simeq L^2-\frac{a^2 E^2}{2}\left(1-\frac{L_z^2}{L^2}\right).
	\end{eqnarray}
	We can rewrite the last equation as follows
	\begin{eqnarray}
		\eta+\xi^2 \simeq \frac{L^2}{E^2}-\frac{a^2 }{2}\left(1-\frac{L_z^2}{L^2}\right).\label{xi1}
	\end{eqnarray}

\begin{figure*}[ht!]
		\centering
	\includegraphics[scale=0.64]{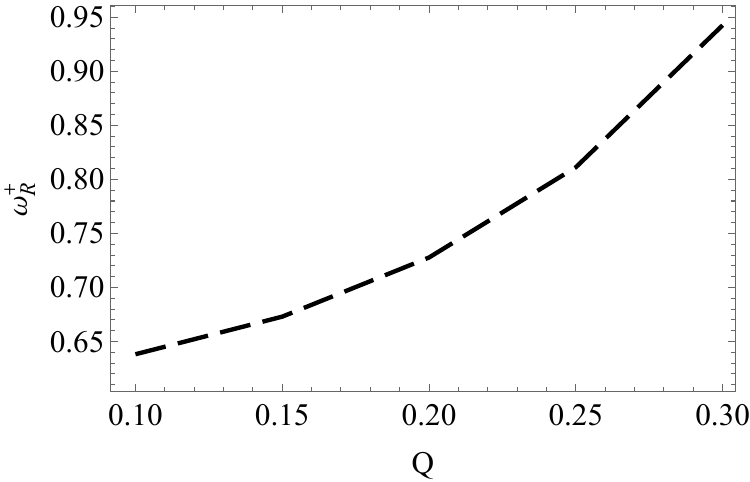}
	\includegraphics[scale=0.64]{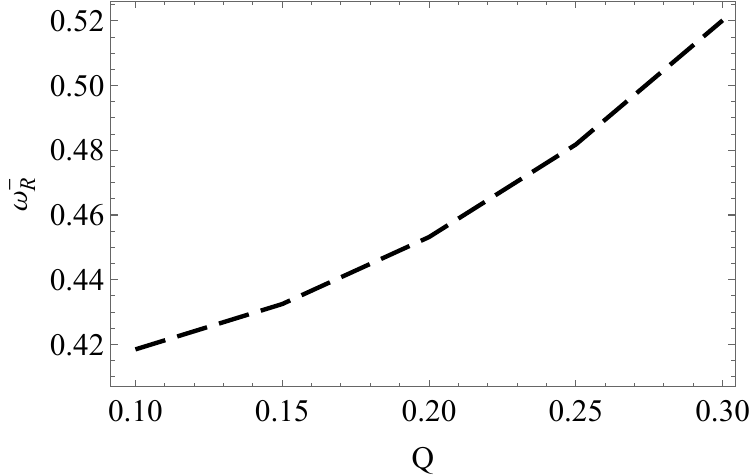}
	\caption{(Left panel) The plot of the real part of the equatorial mode $\omega_R^{+}$ as a function of $Q$. We have fixed $l_0=0.1$  and $M=1$, $a=0.5$ and $l=2$. (Right panel) The plot of the real part of the equatorial mode $\omega_R^{-}$ as a function of $Q$. We have fixed $l_0=0.1$, $M=1$, $a=0.5$ and $l=2$, respectively.   }
	\end{figure*}

 \begin{figure*}[ht!]
		\centering
	\includegraphics[scale=0.65]{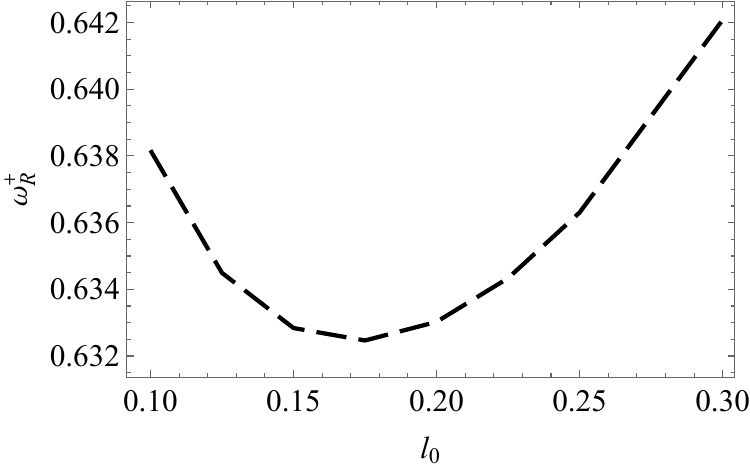}
	\includegraphics[scale=0.65]{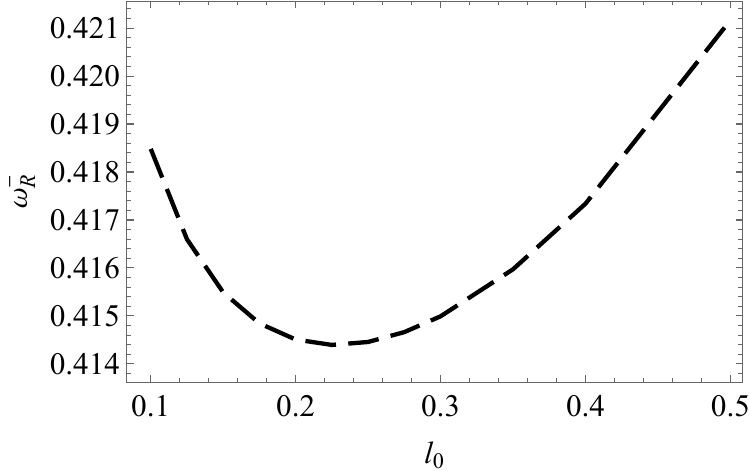}
	\caption{(Left panel) The plot of the real part of the equatorial mode $\omega_R^{+}$ as a function of $l_0$. We have fixed $Q=0.1$  and $M=1$, $a=0.5$ and $l=2$. (Right panel) The plot of the real part of the equatorial mode $\omega_R^{-}$ as a function of $l_0$. We have fixed $Q=0.1$, $M=1$, $a=0.5$ and $l=2$, respectively.   }
	\end{figure*}

  Now, we use the geometric-optics correspondence between the parameters of a QNMs and the conserved quantities along geodesics. In particular we write: \cite{Yang:2012he} 
	\begin{eqnarray}
		L_z  & \Longleftrightarrow & m\\ 
		E & \Longleftrightarrow & \omega_{\Re}\\
		L & \Longleftrightarrow & l+\frac{1}{2}
	\end{eqnarray}
	In the eikonal limit with $m=l \gg 1$, the real part of QNMs is expressed as $\omega_{\Re}=L \Omega_R$. Under this condition, if we introduce $\mu=m/(l+1/2)=1$, the relation can be established using Eq. \eqref{lt} with a positive sign before the orbital frequency:
	\begin{eqnarray}
		\Omega_{prec}=\Omega_{\phi}-\Omega_{\theta},
	\end{eqnarray}
	hence one can obtain
	\begin{eqnarray}
  \Omega_R=\Omega_{\theta}+\Omega_{prec}=\Omega_{\phi},
	\end{eqnarray}
where the term $\Omega_{\theta}$ cancels out. It means that the real part of QNMs is related to the Kepler frequency given by 
	\begin{eqnarray}
		\omega^{\pm}_{\Re}=(l+\frac{1}{2})\frac{-\partial _r g_{t \phi }\pm \sqrt{\left(\partial _r g_{t \phi }\right)^2-\partial _r g_{t t} \partial _r g_{\phi  \phi }}}{\partial _r g_{\phi
				\phi }}\label{rpart}
	\end{eqnarray}
	which is valid in the limit $m=l>>1$. Alternatively, we can consider the mode $m=-l$, which follows that $\mu=-1$, where we can use Eq. \eqref{lt} with a negative sign before the Kelperian frequency, and we arrive again at Eq. \eqref{rpart}. In this case, we can use the following definition to specify the shadow radius \cite{Feng, Jusufi:2020dhz, Jusufi:2022tcw,J1}
	\begin{eqnarray}
		R_{sh}:=\frac{1}{2}\left(\xi^+(r_{0}^+)-\xi^-(r_{0}^-)\right),\label{de1}
	\end{eqnarray}
	provided $\eta(r_{0}^{\pm})=0$. From Eq. \eqref{xi1} one can obtain 
	\begin{eqnarray}
		\xi^{\pm}=\pm \sqrt{\frac{(l+\frac{1}{2})^2}{\omega^2_{\Re}(r_{0}^{\pm})}-\frac{a^2}{2}(1-\mu^2)}.
	\end{eqnarray}
	The correspondence is  precise if we consider the eikonal limit, that is, if we set $\mu=\pm 1$ (namely $ [(m=\pm l)]$, yielding \cite{Jusufi:2020dhz, Jusufi:2022tcw}
	\begin{equation}
		R_{sh}(\mu=\pm 1)=\frac{l+\frac{1}{2}}{2}\left(\frac{1}{\omega_{\Re}(r_{0}^+)}+\frac{1}{\omega_{\Re}(r_{0}^-)}\right).
	\end{equation}

	The final equation aligns with the expression derived in \cite{J1}. Utilizing Eq. \eqref{rpart} and incorporating the metric functions in the equatorial plane, we can derive the following expression for the typical shadow radius:
 \cite{Jusufi:2020dhz}
	\begin{equation}
		R_{sh}=\frac{\sqrt{2}}{2}\left(\sqrt{\frac{ r_0^{+}}{f'(r)|_{r_0^{+}}}}+\sqrt{\frac{ r_0^{-}}{f'(r)|_{r_0^{-}}}}\right).\label{rs}
	\end{equation}
	The preceding equation corresponds precisely to the result established in Ref. \cite{J1}, where the points $r_0^{\pm}$ were determined by solving the equation:
	\begin{equation}
		r_0^2-\frac{2 r_0}{f'(r)|_{r_0^{\pm}}}f(r_0)\mp 2 a \sqrt{\frac{2 r_0}{f'(r)|_{r_0^{\pm}}}}=0.
	\end{equation}
	
	As we elaborated above, in the eikonal limit, we can use Eq. \eqref{rpart}, which can be further simplified as follows 
	\begin{eqnarray}
		\omega_{\Re}^{\pm}=(l+\frac{1}{2}) \frac{1}{a \pm \sqrt{\frac{ 2 r_0^{\pm}}{f'(r)|_{r_0^{\pm}}}} }.
	\end{eqnarray}
	
	Using the last equation, in Fig. 3, we plotted our result for the real part of equatorial QNMs. As we can see, with the increase of charge, the value of $\omega_{\Re}$ increases. This confirms the inverse relation between QNMs and shadow radius, where we found that shadow radius decreases with the increase of the electric charge (see Fig. 2). Furthermore, when varying $l_0$, we observe a reflecting point as initially, $\omega_{\Re}$ decreases with the increase of $l_0$, reaching some minimal value and then increasing faster by increasing $l_0$ (see Fig. 4, right panel). To conclude, it was shown that the reflecting point was resent in the case of shadow radius and QNMs; hence, the correspondence between QNMs and shadow radius holds.  \\

	\subsection{Viewing angle:  $\theta_0=0$ \&  $\theta_0=\pi$}
	We will now focus on the polar orbit, in that case we have to use $\theta=0$. In this specific scenario, it is natural to calculate the shadow radius with viewing angles for the observer, denoted as $\theta_0=0$ and $\theta_0=\pi$. Utilizing celestial coordinates, we can derive:
	\begin{eqnarray}
		X^2+Y^2=\eta+\xi^2+a^2 \cos^2\theta_0.
	\end{eqnarray}
	This shows that, when considering a viewing angle $\theta_0=0$ (or $\theta_0=\pi$), the shadow retains circular disk shape. Therefore, it is natural to employ the following definition \cite{Feng}
	\begin{eqnarray}	R_{sh}:=\sqrt{a^2+\eta(r_0)},\,\,\,\text{with}\,\,\xi(r_0)=0.
	\end{eqnarray}
	In the context of the polar orbit, the azimuthal angular momentum is zero, i.e., $L_z=0$. Under these conditions, and considering the requirements for the existence of circular geodesics, specifically $\dot{r}^2$, we deduce from the radial geodesics:
	\begin{equation}
		(r^2+a^2)^2-[r^2f(r)+a^2]R_{sh}^2=0,\label{def2}
	\end{equation}
	and
	\begin{equation}
		4 r (r^2+a^2)-2 r f(r)R_{sh}^2-r^2f'(r)R_{sh}^2=0,
	\end{equation}
where the definition $R_{sh}^2=\mathcal{K}/E^2+a^2$ (see in particular \cite{Dolan}) has been used. By utilizing Eq. \eqref{def2}, the typical shadow radius can be defined as $R_{sh}\equiv(R_{sh}^+-R_{sh}^-)/2$. Consequently, we obtain:
	\begin{equation}
		R_{sh}=\frac{a^2+r^2}{\sqrt{r^2 f(r)+a^2}}|_{r_0},\label{eq61}
	\end{equation}
	In this context, $r_0$ can be determined by solving the algebraic equation:
	\begin{eqnarray}
		(a^2+r_0^2)^2-\frac{4 [r_0^2 f(r_0)+a^2](a^2+r_0^2)}{r_0 f'(r_0)+2 f(r_0)}=0.
	\end{eqnarray}
	On the other hand, using Eq. (34) and the geometric-optic correspondence, we get the relation between the shadow radius and the real part of QNMs in terms of
	\begin{eqnarray}
		R_{sh}=\sqrt{\frac{(l+1/2)^2}{\omega^2_{\Re}(r_0)}+\frac{a^2}{2}}.
	\end{eqnarray}

 \begin{figure*}[ht!]
		\centering
	\includegraphics[scale=0.65]{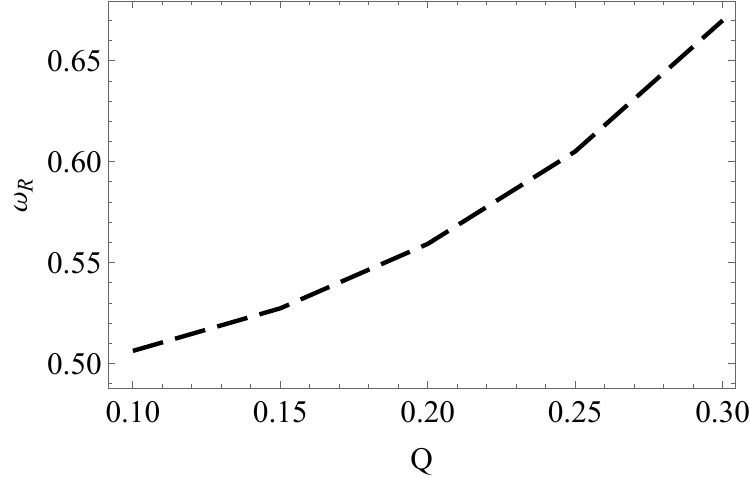}
	\includegraphics[scale=0.65]{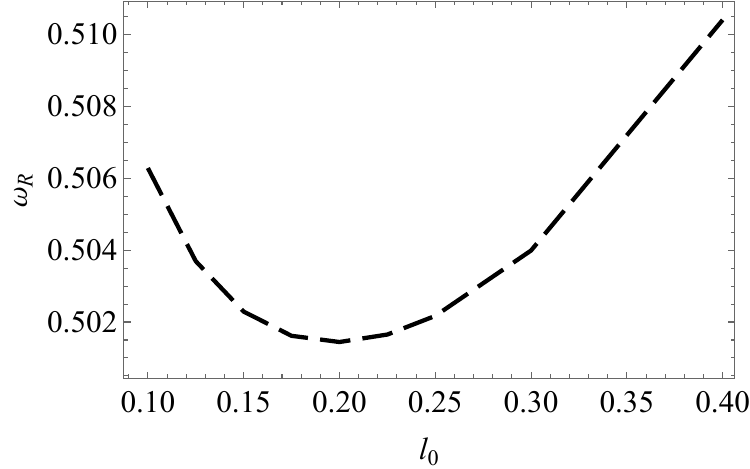}
	\caption{(Left panel) The plot of the real part of the polar mode $\omega_R$ as a function of $Q$. We have fixed $l_0=0.1$  and $M=1$, $a=0.5$ and $l=2$. (Right panel) The plot of the real part of the polar mode $\omega_R$ as a function of $l_0$. We have fixed $Q=0.1$ and $M=1$, along with $a=0.5$ and $l=2$.   }
	\end{figure*}
 
	Finally, using Eq. (51) along with the last equation, we get for the real part of QNMs a compact equation
	\begin{equation}
		\omega_{\Re}=\left(l+\frac{1}{2}\right)\sqrt{\frac{2 (r^2f(r)+a^2)\zeta}{\zeta}},
	\end{equation}
	where $\zeta=2r^4+4a^2r^2+a^4-a^2r^2 f(r)$. In Fig. 5 (left panel), we plotted our result for the real part of polar QNMs. As we can see, with the increase of charge, the value of $\omega_R$ increases. On the other hand, for the polar QNMs, when varying $l_0$ we observe a reflecting point, that is, initially $\omega_{\Re}$ decreases with the increase of $l_0$, reaching some minimal value and then increasing faster by increasing $l_0$ (see Fig. 5, right panel). It should be noted that for larger values of $l$, we should obtain more accurate values for QNMs, as the correspondence is accurate in the eikonal limit.

\section{Superradiance}
In this section, we intend to study the superradiance scattering of a massive scalar field $\Phi$ off a rotating black hole in T-duality. To this end, the Klein-Gordon equation in curved
spacetime is taken into consideration. It is given by
\begin{eqnarray}\notag
\left(\bigtriangledown_{\alpha}\bigtriangledown^{\alpha}+\mu^{2}\right)\Phi(t,r,\theta,\phi)=0,
\label{KG}
\end{eqnarray}
which can be also written as 
\begin{eqnarray}\notag
\left[\frac{-1}{\sqrt{-g}}\partial_{\sigma}\left(g^{\sigma
\tau}\sqrt{-g}\partial_{\tau}\right)+\mu^{2}\right]\Phi(t,r,\theta,\phi)=0.
\end{eqnarray}
Here, $\mu$  represents the mass of the scalar field $\Phi$. With the help of the standard separation of variables method and using the following ansatz with the standard
Boyer-Lindquist coordinates $(t, r, \theta, \phi)$,
\begin{eqnarray}
\Phi(t, r, \theta, \phi)=R_{\omega j m}(r) \Theta(\theta) e^{-i
\omega t} e^{i m \phi}, \label{PHI}
\end{eqnarray}
with $\quad j \geq 0, \quad-j \leq m \leq j,
\quad \omega>0$. We will be able to separate Eq. (\ref{KG}) into radial and angular parts. In Eq. (\ref{PHI}), $R_{\omega j m}(r)$ and $\Theta(\theta)$ are the radial function and oblate spheroidal wave function, respectively. The symbols $j$, $m$, and $\omega$ represent the angular
eigenfunction, angular quantum number, and the positive frequency
of the field  under investigation as measured by a far-away
observer, respectively. The Eq. (\ref{KG}), with the help of the ansatz (\ref{PHI}), yields following radial and angular equations:   
\begin{eqnarray}
&&\frac{d}{d r}(\Delta \frac{d R_{\omega j m}(r)}{d
r})+(\frac{((r^2+a^{2} \omega-am
)^{2}}{\Delta})R_{\omega l m}(r)
 \nonumber \\
&&-(\mu^{2} r^2+j(j+1)+a^{2} \omega^{2}-2 m \omega
a) R_{\omega l m}(r)=0, \label{RE}
\end{eqnarray}
\begin{eqnarray}\notag
&&\sin \theta \frac{d}{d \theta}\left(\sin \theta \frac{d
\Theta_{\omega j m}(\theta)}{d \theta}\right)+\\\notag
&+&\left(j(j+1) \sin
^{2} \theta-\left(\left(a \omega \sin ^{2}
\theta-m\right)^{2}\right)\right)\Theta_{\omega j m}(\theta)\nonumber \\
&+& a^{2} \mu^{2} \sin ^{2} \theta \cos ^{2} \theta~
\Theta_{\omega j m}(\theta)=0.
\end{eqnarray}
\subsection{Amplification factor for superradiance}
Since our intention is to investigate the scattering of $\Phi$, we will only consider Eq. (\ref{RE}) for the rest of the paper. Following articles \cite{vb:2014, gv:2016}, the general solution of the radial equation (\ref{RE}) can be obtained. Next, We introduce a Regge-Wheeler-like
coordinate $r_{*}$ defined by
\begin{eqnarray}
r_{*} \equiv \int d r \frac{r^2+a^{2}}{\Delta},
\end{eqnarray}
with $r_{*} \rightarrow-\infty$ at event horizon and $ r_{*} \rightarrow \infty $ at infinity.  We then define a new radial function $\mathcal{P}_{\omega jm}\left(r_{*}\right)=\sqrt{r^2+a^{2}} R_{\omega j m}(r)$ to transform the radial equation (\ref{RE}) into the desired shape. A few steps of algebra provide the following equation: 
\begin{equation}
\frac{d^{2} \mathcal{P}_{\omega l m}\left(r_{*}\right)}{d
r_{*}^{2}}+V_{\omega j m}(r) \mathcal{P}_{\omega j
m}\left(r_{*}\right)=0, \label{RE1}
\end{equation}
where $V_{\omega j m}(r)$ is the effective potential given by
\begin{eqnarray}\nonumber
V_{\omega j m}(r)&=&\left(\omega-\frac{m
a}{r^2+a^{2}}\right)^{2}-\frac{\Delta}{\left(r^2
+a^{2}\right)^{2}}\Big[ j(j+1)\\\notag
&+& a^{2}
\omega^{2}-2 m a \omega+\mu^{2}r^2\\
&+&\sqrt{r^2+a^{2}}\frac{d}{dr}\left(\frac{r\Delta
}{\left(r^2+a^{2}\right)^{\frac{3}{2}}}\right)\Big],
\label{POT}
\end{eqnarray}
So, the resulting effect is that we now study the scattering of the scalar field $\Phi$ under the
effective potential (\ref{POT}). The asymptotic behavior of the potential (\ref{POT}) at the event
horizon and at spatial infinity play a crucial role in obtaining the asymptotic solutions of the radial equation (\ref{RE}). The limiting value of the potential at the event horizon is 
\begin{eqnarray}
\lim _{r \rightarrow r_{eh}} V_{\omega j
m}(r)=\left(\omega-m \Omega_{t}\right)^{2}
\equiv k_{e h}^{2},\quad \Omega_{t}=\frac{a}{r_{eh}^2+a^2},
\end{eqnarray}
and the limiting value at spatial infinity is
\begin{equation}
\lim _{r \rightarrow \infty} V_{\omega j m}(r)=\omega^{2}-\mu^{2}\equiv k_{\infty}^{2}.
\end{equation}
The potential has distinct, constant limiting values at the extreme points. With the asymptotic values of the potential at our disposal, we are now able to obtain the asymptotic radial solutions. After a few steps of algebra, we obtain 
\begin{equation}\label{AS}
R_{\omega j m}(r) \rightarrow\left\{\begin{array}{cl}
\frac{\mathcal{N}_{i n}^{eh} e^{-i k_{eh} r_{*}}}{\sqrt{r_{e h}^2
+a^{2}}} & \text { for } r \rightarrow r_{e h} \\
\mathcal{N}_{i n}^{\infty} \frac{e^{-i k_{\infty}
r_{*}}}{r}+\mathcal{N}_{r e f}^{\infty} \frac{e^{i k_{\infty}
r_{*}}}{r} & \text { for } r \rightarrow \infty
\end{array}\right.
\end{equation}
Here, $\mathcal{N}_{in}^{eh}$ and $\mathcal{N}_{in}^{\infty}$ correspond to the amplitude of the
incoming scalar wave at the event horizon($r_{eh}$) and infinity $(\infty)$, respectively. $\mathcal{N}_{ref}^{\infty}$ represents the amplitude of the
reflected part of the scalar wave at infinity $(\infty)$. Matching Wronskinas at the event horizon and infinity, we obtain
\begin{equation}
\left|\mathcal{N}_{r e f}^{\infty}\right|^{2}=\left|\mathcal{N}_{i
n}^{\infty}\right|^{2}-\frac{k_{e
h}}{k_{\infty}}\left|\mathcal{N}_{i n}^{e h}\right|^{2}.
\label{AMP}
\end{equation}

\begin{figure*}[ht!]
		\centering
	\includegraphics[scale=0.70]{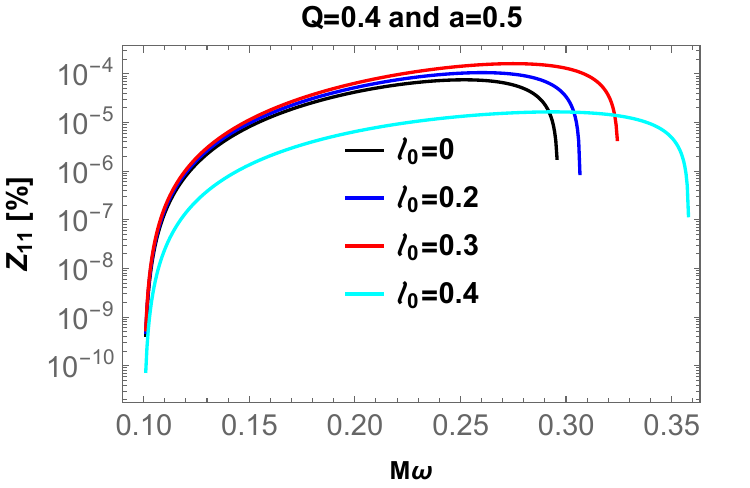}
	\includegraphics[scale=0.70]{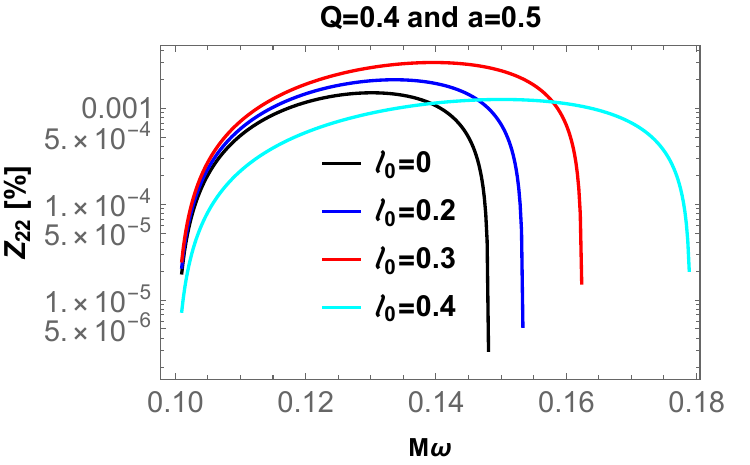}
	\caption{Variation of percentage amplification factors against $\omega$ for
different values of $l_0$ with $Q=0.4$. }\label{l}
	\end{figure*}

The above relation shows that the superradiance scattering of the scalar field is possible when $frac{k_{eh}}{k_{\infty}}<0$, i.e., $\omega<m \Omega_{T}$. To obtain the analytical expressions of amplitudes, we employ the asymptotic matching method proposed in \cite{aa:1973a, aa:1973b}. In this procedure, we obtain separate solutions for the two overlapping regions: one is the near region where $r-r_{eh}\ll \omega$ and another is the far region where $r-r_{eh} >> 1$. We then expand the resulting solutions to match them in the overlapping region which finally yields the desired expressions of amplitudes. Following the procedure and assumptions given in \cite{skj:2022, skj:2023a, skj:2023b}, we get the amplification factor as
\begin{equation}
Z_{j m} \equiv \frac{\left|\mathcal{N}_{r e
f}^{\infty}\right|^{2}}{\left|\mathcal{N}_{i
n}^{\infty}\right|^{2}}-1, \label{AMPZ}
\end{equation}
where
\begin{widetext}
\begin{eqnarray}
\mathcal{N}_{in}^{\infty}&=&\frac{b(-2
i)^{-\frac{1+\sqrt{1+4j(j+1)}}{2}}}{\sqrt{(\omega^{2}-\mu^{2})}} \cdot
\frac{\Gamma(\sqrt{1+4j(j+1)}) \Gamma(1+\sqrt{1+4j(j+1)})}{\Gamma\left(\frac{1+\sqrt{1+4j(j+1)}}{2}-2
iB\right)\left(\Gamma\left(\frac{1+\sqrt{1+4j(j+1)}}{2}\right)\right)^{2}}\times
\\\nonumber &&\Gamma(1-2 iB) k_{t}^{\frac{1-\sqrt{1+4j(j+1)}}{2}}+\frac{b(-2
i)^{\frac{\sqrt{1+4j(j+1)}-1}{2}}}{\sqrt{(\omega^{2}-\mu^{2})}}
\times \\\nonumber &&\frac{\Gamma(1-\sqrt{1+4j(j+1)})
\Gamma(-\sqrt{1+4j(j+1)})}{\left(\Gamma\left(\frac{1-\sqrt{1+4j(j+1)}}{2}\right)\right)^{2}
\Gamma\left(\frac{1-\sqrt{1+4j(j+1)}}{2}-2 iB\right)} \Gamma(1-2 i B)
k_{t}^{\frac{1+\sqrt{1+4j(j+1)}}{2}},
\end{eqnarray}
and
\begin{eqnarray}
\mathcal{N}_{ref}^{\infty}&=&\frac{b(2
i)^{-\frac{1+\sqrt{1+4j(j+1)}}{2}}}{\sqrt{(\omega^{2}-\mu^{2})}} \cdot
\frac{\Gamma(\sqrt{1+4j(j+1)}) \Gamma(1+\sqrt{1+4j(j+1)})}{\Gamma\left(\frac{1+\sqrt{1+4j(j+1)}}{2}-2
iB\right)\left(\Gamma\left(\frac{1+\sqrt{1+4j(j+1)}}{2}\right)\right)^{2}}\times
\\\nonumber &&\Gamma(1-2 iB) k_{t}^{\frac{1-\sqrt{1+4j(j+1)}}{2}}+\frac{b(2
i)^{\frac{\sqrt{1+4j(j+1)}-1}{2}}}{\sqrt{(\omega^{2}-\mu^{2})}}
\times \\\nonumber &&\frac{\Gamma(1-\sqrt{1+4j(j+1)})
\Gamma(-\sqrt{1+4j(j+1)})}{\left(\Gamma\left(\frac{1-\sqrt{1+4j(j+1)}}{2}\right)\right)^{2}
\Gamma\left(\frac{1-\sqrt{1+4j(j+1)}}{2}-2 iB\right)} \Gamma(1-2 i B)
k_{t}^{\frac{1+\sqrt{1+4j(j+1)}}{2}}.
\end{eqnarray}
\end{widetext}
Here, $B=\frac{\omega-m\Omega_{t}}{r_{eh}-r_{ch}}r_{eh}^2$ and $k_t=\left(r_{eh}-r_{ch}\right)\sqrt{\omega^2-\mu^2}$, $r_{ch}$ being the Cauchy horizon. As evident from the Eq. (\ref{AMPZ}), we have superradiance when $Z_{jm}>0$. Since cases $m<0$ are non-superradiant, we ignore them here as we are interested in the occurrence of superradiance. We intend to study the effect of the "Zero-point length" $l_0$ and the charge Q. We have taken $M=1$, $\mu=0.1$, and $a=0.5$.

 \begin{figure*}[ht!]
		\centering
	\includegraphics[scale=0.67]{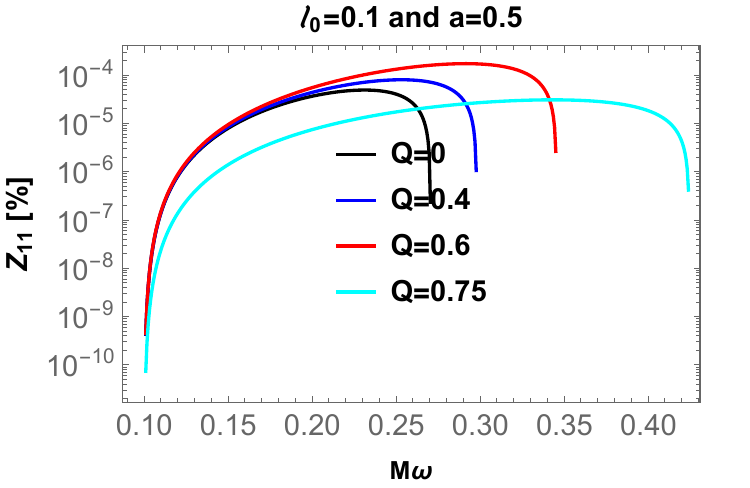}
	\includegraphics[scale=0.67]{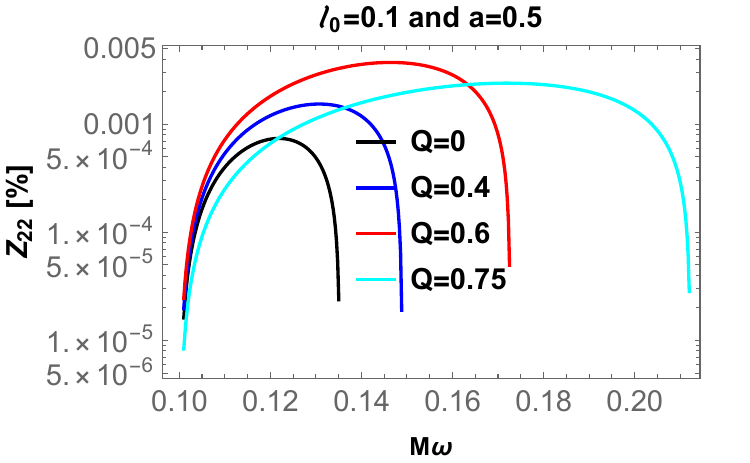}
	\caption{Variation of percentage amplification factors against $\omega$ for
different values of $Q$ with $l_0=0.1$. }\label{Q}
	\end{figure*}

\begin{figure*}[ht!]
		\centering
	\includegraphics[scale=0.50]{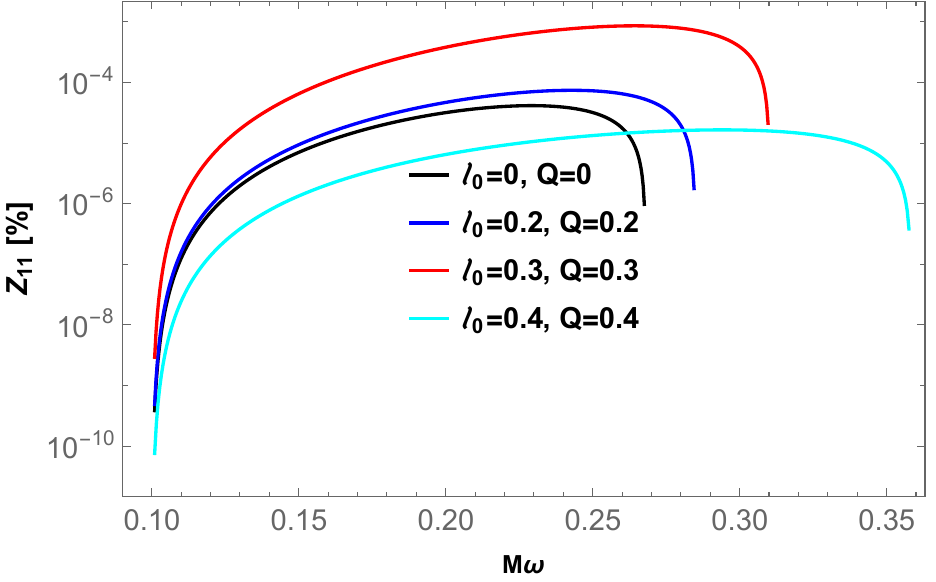}
	\includegraphics[scale=0.50]{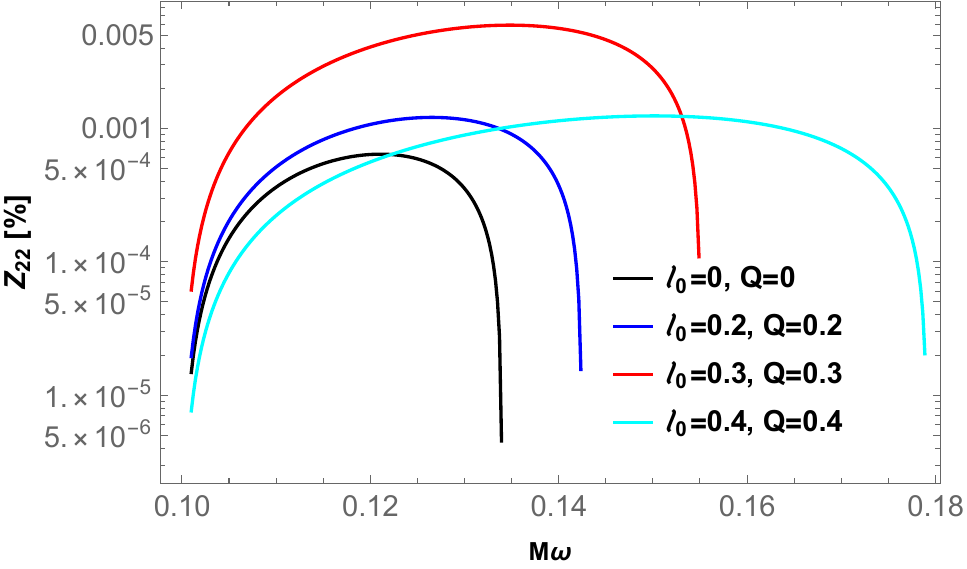}
	\caption{Comparison between Kerr black hole and the rotating black hole in T-duality.} \label{compare}
	\end{figure*}

\begin{figure*}[ht!]
		\centering
	\includegraphics[scale=0.60]{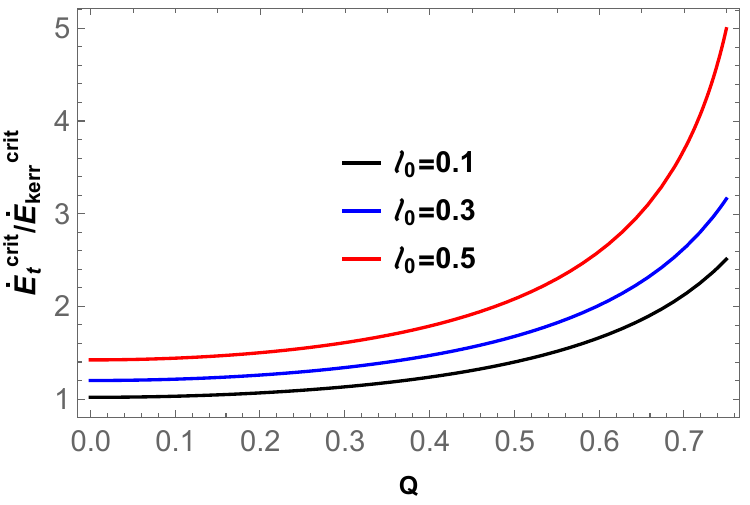}
	\caption{Ratio of the outgoing energy fluxes as a function of charge Q for different values of $l_0$.}
\label{flux}
	\end{figure*}

Through the display of plots (\ref{l}, \ref{Q}), we illustrate the effect of parameters $l_0$ and Q on the superradiance scattering. Both the plots show that the amplification factor initially increases with an increase in $l_0$ and Q and then starts decreasing. This trend is also visible in Fig. (\ref{compare}), where we make a comparison between the 
Kerr black hole and the rotating black hole in T-duality. It is also observed that the black hole in T-duality allows wider range of frequency to be amplified. For those frequencies that lie between the critical frequencies of the Kerr black hole and black hole in T-duality, waves get absorbed by the Kerr black hole but superradiantly scattered by the black hole in T-duality. This has larger impact on the luminosity. To illustrate this fact further, we consider a monochromatic massive scalar field. The outgoing energy flux of the field, measured by an observer at infinity, is given by \cite{rvp:2015}
\begin{equation}
    \dot{E}=\frac{\omega k_{\infty}}{2}\left|\mathcal{N}_{r ef}^{\infty}\right|^{2}= \frac{\omega k_{\infty}}{2} \frac{\left|\mathcal{N}_{r e
f}^{\infty}\right|^{2}}{\left|\mathcal{N}_{i
n}^{\infty}\right|^{2}}\left|\mathcal{N}_{i
n}^{\infty}\right|^{2}
\end{equation}
We take the ratio of energy fluxes for the Kerr black hole and the black hole in T-duality at the respective critical superradiant frequency to make a meaningful comparison. The ratio of critical energy fluxes for massless scalar fields with same $\left|\mathcal{N}_{in}^{\infty}\right|$  is 
\begin{equation}
   \frac{\dot{E}^{crit}_{T}}{\dot{E}^{crit}_{Kerr}}=\frac{\Omega_{t}^2}{\Omega_{kerr}^2}>1
\end{equation}
The ratio being greater than unity shows that the black hole in T-duality would be brighter than the Kerr black hole. To illustrate the dependency of the ratio on the parameters $l_0$ and Q, we plot the ratio as a function of Q for different values of $l_0$ in Fig. (\ref{flux}). Here, the spin parameter is fixed at $0.5$.

\subsection{Superradiant instability for the rotating black hole in T-duality}

In this section, we investigate the superradiant instability of the combined system of rotating black hole and scalar field due to the phenomenon known as black hole bomb \cite{sa:1974, wh:1972}. In this phenomenon, the superradiant field gets trapped by a potential well and gets repeatedly amplified, resulting in exponential growth of the amplitude and hence superradiance instability.

From equation (\ref{RE}), we have
\begin{eqnarray}
\Delta \frac{d}{d r}\left(\Delta \frac{d R_{\omega j m}}{d
r}\right)+\mathcal{K} R_{\omega j m}=0, \label{MRE}
\end{eqnarray}
where for a slowly rotating black hole $(\hat{a} \omega \ll 1)$
$$
\mathcal{K} \equiv \left(\left(r^2+a^{2}\right) \omega-ma\right)^{2}+\Delta\left(2 ma
\omega-j(j+1)-\mu^{2}r^2\right).
$$

Following the black hole bomb mechanism, solutions for the radial equation (\ref{MRE}) is given by
$$
R_{\omega j m} \sim\left\{\begin{array}{ll}
e^{-i(\omega-m \Omega_{t}) r_{*}} & \text { as } r \rightarrow r_{e h}
\left(r_{*} \rightarrow-\infty\right) \\
\frac{e^{-\sqrt{\mu^{2}-\omega^{2}} r_{*}}}{r} & \text { as } r
\rightarrow \infty  \left(r_{*} \rightarrow \infty\right)
\end{array}\right.
$$
We now introduce a new radial function:
$$
\psi_{\omega j m} \equiv \sqrt{\Delta} R_{\omega j m},
$$
that converts the radial equation (\ref{MRE}) into
$$
\left(\frac{d^{2}}{d r^{2}}+\omega^{2}-V\right)
\psi_{\omega j m}=0.
$$
with
\begin{figure*}[ht!]
		\centering
	\includegraphics[scale=0.45]{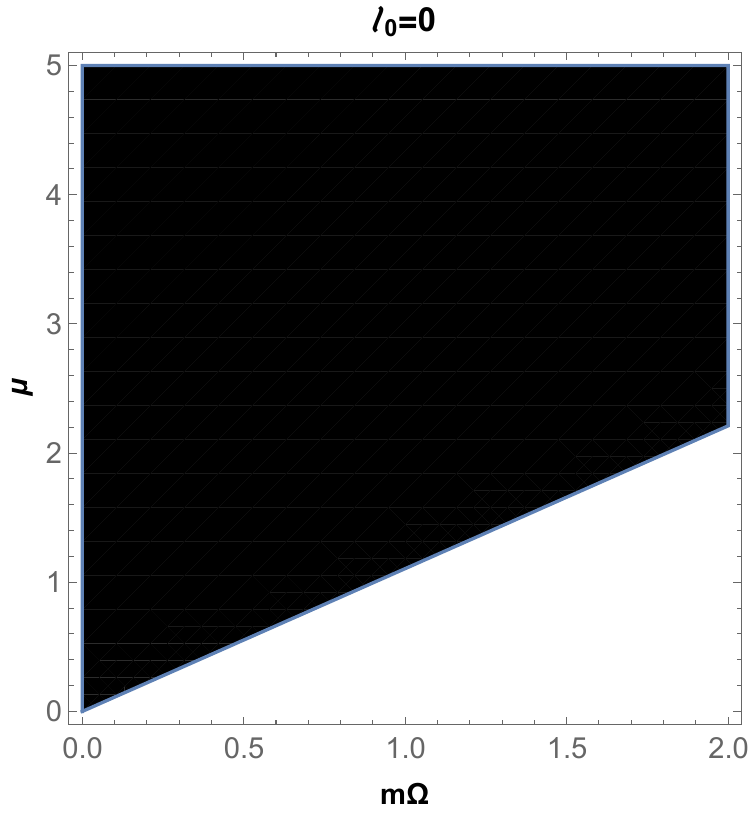}
	\includegraphics[scale=0.45]{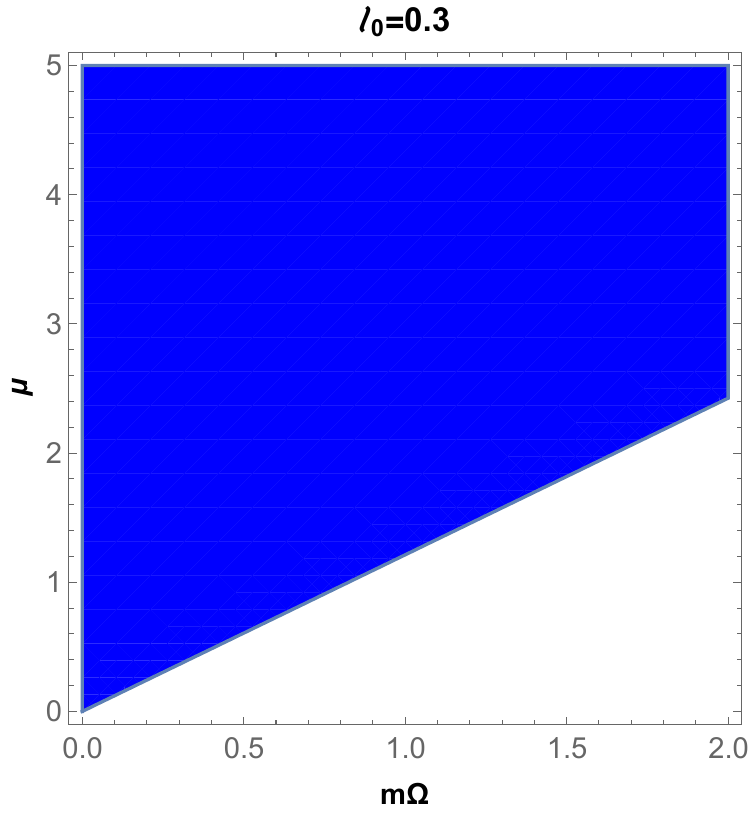}
         \includegraphics[scale=0.45]{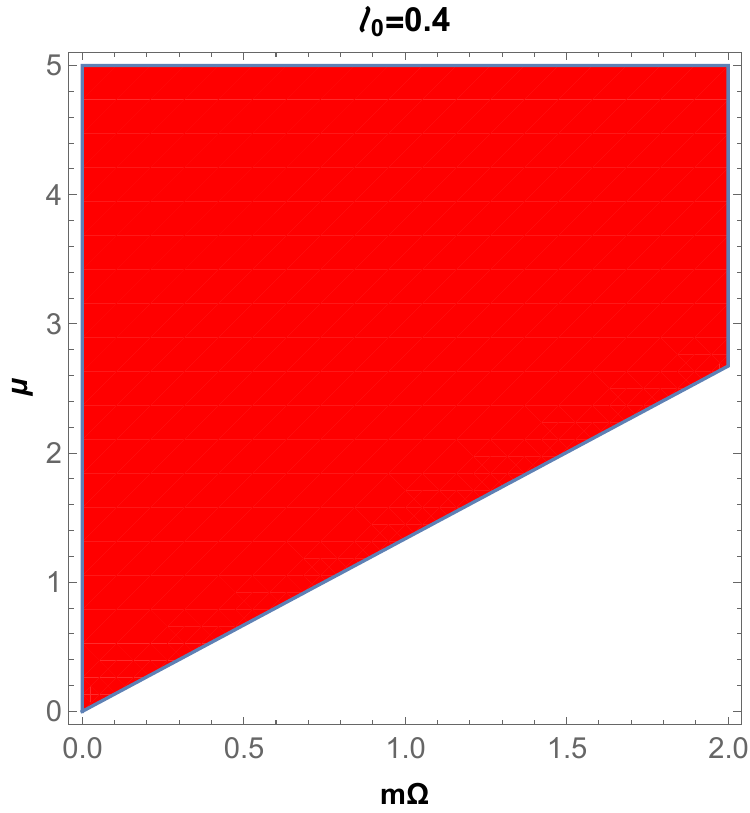}
	\caption{Parameter space($m\Omega$-$\mu$) for massive scalar field
where colored area represents region with stable dynamics and
non-colored area represents region with unstable dynamics. Here $Q=0.4$.}
\label{spacel}
	\end{figure*}

 \begin{figure*}[ht!]
		\centering
	\includegraphics[scale=0.45]{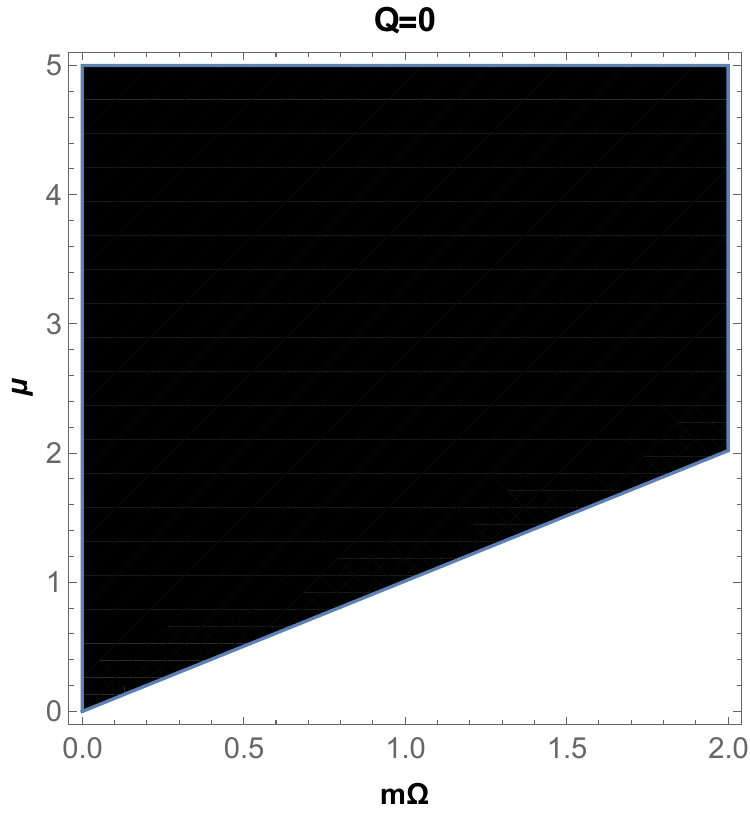}
	\includegraphics[scale=0.45]{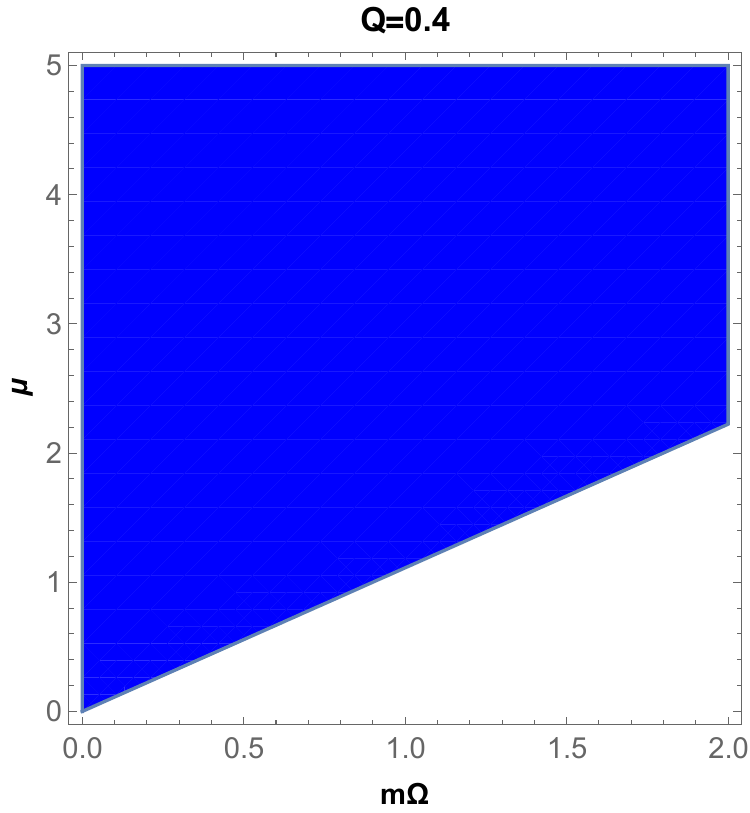}
         \includegraphics[scale=0.45]{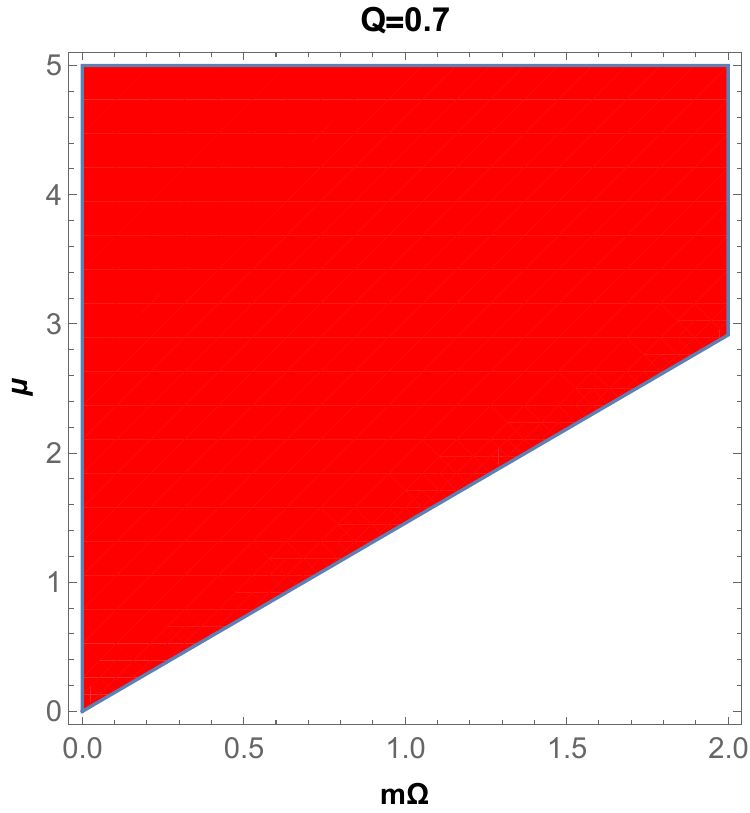}
	\caption{Parameter space($m\Omega$-$\mu$) for massive scalar field
where colored area represents region with stable dynamics and
non-colored area represents region with unstable dynamics. Here $l_0=0.1$.}
\label{spaceQ}
	\end{figure*}

$$
\omega^{2}-V=\frac{\mathcal{K}+\frac{1}{4}(\frac{d\Delta}{dr})^2-\frac{\Delta}{2}\frac{d^2\Delta}{dr^2}}{\Delta^{2}},
$$
which is the Regge-Wheel equation. Retaining terms up to the order
$\mathcal{O}\left(1 / r\right)$, we obtain the asymptotic form of the
effective potential $V(r)$ as
$$
V(r)=\mu^{2}-\frac{4 M
\omega^{2}}{r}+\frac{2 M \mu^{2}}{r}.
$$
We impose the condition $V^{\prime} \rightarrow 0^{+}$ as $r \rightarrow
\infty$ essential for the realization of black hole bomb mechanism \cite{sh:2012}. Taking into consideration the fact that the superradiance
amplification of scattered waves occurs when $\omega<m
\Omega_{t}$, we obtain
$$
\frac{\mu}{\sqrt{2}}<\omega^{2}<m \Omega_{t},
$$
in which the integrated system of the rotating black hole in T-duality and the massive scalar field may experience a superradiant instability,
known as the black hole bomb. The dynamics of the massive
scalar field in black hole will remain stable when
$\mu\geq\sqrt{2}m\Omega_{t}$. Plots (\ref{spacel}, \ref{spaceQ}) show that the parameter space $(m\Omega-\mu)$ over which the combined system of black hole and scalar field remains stable decreases with an increase in either $l_0$ or Q.

\section{Final remarks}
In this current study, we have explored the shadow images, the correlation between QNMs and the shadow radius, as well as the investigation of the superradiance effect and the stability of charged black holes within the context of T-duality. Our findings indicate that the shadow radius decreases as the electric charge increases. Particularly intriguing is the observation regarding the quantum deformed parameter $l_0$, where a distinct phenomenon emerges—a reflecting point is identified. At this point, the shadow radius initially increases with an increase in $l_0$ and subsequently decreases.
Furthermore, our exploration extended to the analysis of equatorial and polar QNMs, revealing a notable pattern: the real part of QNMs amplifies with an increase in charge. Nevertheless, an intriguing turning point is uncovered when varying $l_0$. Consequently, this shows the inverse relationship between QNMs and shadow radius in our investigation. Broadly speaking, in the context of astrophysical black holes, the predominant influence is attributed to the electric charge, while the role of $l_0$ is expected to be important only for smaller black holes. It it still an open question if such quantum effects can have an impact for large black holes. 

For the superradiance the amplification factor shows the existence of reflecting points for both the parameters. The critical frequency, however, increases with an increase in the electric charge and the quantum deformed parameter. This has a direct bearing on the luminosity of the black hole as observed at infinity. A wider range of allowed frequency for superradiance makes black holes in T-duality brighter than the Kerr black hole.


\begin{thebibliography}{}
\bibitem{m87} The EHT Collaboration, Astrophys. J. {\bf875}, L1 (2019).
		
		\bibitem{m871} The EHT Collaboration, Astrophys. J. Lett. {\bf910}, L13 (2021).
		
		 \bibitem{EHT2022-1}
K. Akiyama \textit{et al.} [Event Horizon Telescope], Astrophys.J.Lett. 930 (2022) L12

 \bibitem{EHT2022-2}
K. Akiyama \textit{et al.} [Event Horizon Telescope],
Astrophys.J.Lett. 930 (2022) L13

 \bibitem{EHT2022-3} 
K. Akiyama \textit{et al.} [Event Horizon Telescope],
Astrophys.J.Lett. 930 (2022) L14
		
   
\bibitem{2}
K.~Akiyama \textit{et al.} [Event Horizon Telescope],
Astrophys. J. Lett. \textbf{875}, L1 (2019)


 
\bibitem{3}
K.~Akiyama \textit{et al.} [Event Horizon Telescope],
Astrophys. J. Lett. \textbf{875}, no.1, L2 (2019)

 \bibitem{4}
B.~P.~Abbott {\it et al.} [LIGO Scientific and Virgo Collaborations],
Phys.\ Rev.\ Lett.\  {\bf 116}, no. 6, 061102 (2016)


\bibitem{penrose1}R. Penrose, Gen. Relativ. Gravit. 34, 1141 (2002).
\bibitem{penrose2} R. Penrose and R. M. Floyd, Nature (London) 229, 177 (1971).
\bibitem{misner} C.W. Misner, Phys. Rev. Lett. 22, 1071 (1969).
\bibitem{yb1}Ya. B. Zel’dovich, JETP Lett. 14, 180 (1971).
\bibitem{yb2} Ya. B. Zel’dovich, J. Exp. Teor. Phys. 35, 1085 (1972),
\bibitem{hawking1} S.W. Hawking, Phys. Rev. Lett. 26, 1344 (1971).
\bibitem{bekenstein1} J. D. Bekenstein, Phys. Rev. D 7, 949 (1973).
\bibitem{bekenstein2} J. D. Bekenstein and M. Schiffer, Phys. Rev. D 58, 064014 (1998).
\bibitem{teukolsky} S. A. Teukolsky, Astrophys. J. 185, 635 (1973).
\bibitem{hawking2} S.W. Hawking, Commun. Math. Phys. 43, 199 (1975); 46,206(E) (1976).
\bibitem{richartz} M. Richartz, S. Weinfurtner, A. J. Penner, and W. G. Unruh, Phys. Rev. D 80, 124016 (2009).
\bibitem{cardoso} V. Cardoso and P. Pani, Classical Quantum Gravity 30,045011 (2013).
\bibitem{ge} G. Eskin, Rev. Math. Phys. 28, 1650025 (2016).
\bibitem{rv} R. Vicente, V. Cardoso, and J. C. Lopes, Phys. Rev. D 97,084032 (2018).
\bibitem{brito} R. Brito, V. Cardoso, and P. Pani, Lect. Notes Phys. 906, 1(2015).
\bibitem{page}D.N. Page, Phys. Rev. D 13 (1976) 198.
\bibitem{mr} M. Richartz and A. Saa, Phys. Rev. D 88, 044008(2013).
\bibitem{cardoso1} V. Cardoso, R. Brito, and J. L. Rosa, Phys. Rev. D 91,124026 (2015).
\bibitem{kg} K. Glampedakis, S. J. Kapadia, and D. Kennefick, Phys.Rev. D 89, 024007 (2014).
\bibitem{khodadi}A. Rahmani, M. Khodadi, M. Honardoost, and H. R. Sepangi, Nucl. Phys. B960, 115185 (2020).
\bibitem{ads0}Cardoso V., Dias O. J., Phys.Rev. D70 (2004) 084011,
\bibitem{vb:2014} V. B. Bezerra , H. S. Vieira and A. A. Costa: Class.Quantum Grav. 31 045003  (2014)
\bibitem{gv:2016} G. V. Kraniotis: Class.Quant.Grav. 33 225011 (2016)
\bibitem{aa:1973a}A. A. Starobinsky, Sov. Phys. JETP 37, 28 (1973), http://www.jetp.ras.ru/cgi-bin/e/index/e/37/1/p28?a=list.
\bibitem{aa:1973b}A. A. Starobinsky and S.M. Churilov, Zh. Eksp.
Teor. Fiz. 65, 3 (1973), http://www.jetp.ras.ru/cgibin/
e/index/e/38/1/p1?a=list [Sov. Phys. JETP 38, 1
(1973)].
\bibitem{skj:2022} S. Kr. Jha, A. Rahaman Eur. Phys. J. C (2022) 82:728
\bibitem{skj:2023a} S. Kr. Jha et. al. PHYSICAL REVIEW D 107, 084052 (2023)
\bibitem{skj:2023b} S. Kr. Jha, A. Rahaman Physics of the Dark Universe 42 (2023) 101327
\bibitem{rvp:2015}R. Brito, V. Cardoso, and P. Pani, Lect. Notes Phys. 906, 1 (2015).
\bibitem{sa:1974}S.A. Teukolsky, W.H. Press, Astrophys. J. 193 (1974) 443.
\bibitem{wh:1972} W.H. Press, S.A. Teukolsky, Nature 238 (1972) 211–212.
\bibitem{sh:2012}S. Hod, Phys. Lett. B 708 (2012) 320 [arXiv:1205.1872 [gr-qc]].

\bibitem{Gaete:2022une}
P.~Gaete and P.~Nicolini,
Phys. Lett. B \textbf{829} (2022), 137100
\bibitem{Gaete:2022ukm}
P.~Gaete, K.~Jusufi and P.~Nicolini,
Phys. Lett. B \textbf{835} (2022), 137546
\bibitem{Jusufi:2020dhz}
K.~Jusufi,
Phys. Rev. D \textbf{101} (2020) no.12, 124063

\bibitem{Jusufi:2022tcw}
K.~Jusufi, M.~Azreg-A\"\i{}nou, M.~Jamil and Q.~Wu,
Universe \textbf{8} (2022) no.4, 210

		
		
		

		
\bibitem{Shaikh:2019fpu}
R.~Shaikh,
Phys. Rev. D \textbf{100}, 024028 (2019).

	\bibitem{cardoso} V. Cardoso, A. S. Miranda, E. Berti, H. Witek, and V. T. Zanchin, Phys. Rev. D {\bf79}, 064016 (2009).
		
		\bibitem{stefanov}
		I.~Z.~Stefanov, S.~S.~Yazadjiev and G.~G.~Gyulchev,
		Phys.\ Rev.\ Lett.\  {\bf 104}, 251103 (2010).
		
		
		\bibitem{J1} K. Jusufi, Phys. Rev. D {\bf 101}, 124063 (2020).

\bibitem{Jusufi:2019ltj}
K.~Jusufi,
Phys. Rev. D \textbf{101} (2020) no.8, 084055

\bibitem{Konoplya:2017wot}
R.~A.~Konoplya and Z.~Stuchl\'\i{}k,
Phys. Lett. B \textbf{771} (2017), 597-602
		
\bibitem{Yang:2012he}
H.~Yang, D.~A.~Nichols, F.~Zhang, A.~Zimmerman, Z.~Zhang and Y.~Chen,
Phys. Rev. D \textbf{86}, 104006 (2012).
		
		\bibitem{Y} H. Yang, Phys. Rev. D {\bf103}, 084010 (2021).
		

		\bibitem{c1}
		B.~Cuadros-Melgar, R.~D.~B.~Fontana and J.~de Oliveira,
		Phys. Lett. B {\bf811}, 135966  (2020).
	


		\bibitem{mash} B. Mashhoon, Phys. Rev. D {\bf31}, 290 (1985).
		
		
	
		\bibitem{Dolan}
		S.~R.~Dolan,
		Phys. Rev. D \textbf{82}, 104003 (2020).

  
	\bibitem{Feng}
		X.~H.~Feng and H.~Lu,
		arXiv:1911.12368 [gr-qc].
		

	\bibitem{Zhang:2019glo}
M.~Zhang and M.~Guo,
Eur. Phys. J. C \textbf{80}, 790 (2020).

\bibitem{Azreg-Ainou:2014pra}
M.~Azreg-A\"\i{}nou,
Phys. Rev. D \textbf{90} (2014) no.6, 064041

\end{thebibliography}
\end{document}